\documentclass[a4paper,11pt]{article}
\usepackage{pos}
\usepackage{deluxetable}

\title{The Recurrent Nova Population in M31}
\ShortTitle{M31 RNe}

\author*[a]{Allen W. Shafter}
\author[b]{Kamil Hornoch}

\affiliation[a]{San Diego State University,\\
  Department of Astronomy, San Diego, CA, USA}

\affiliation[b]{Astronomical Institute of the Czech Academy of Sciences,\\
Fri\v{c}ova 298, CZ-251 65 Ond\v{r}ejov, Czech Republic}

\emailAdd{ashafter@sdsu.edu}
\def\lessim{\mathrel{\hbox{\rlap{\hbox{\lower4pt\hbox{$\sim$}}}\hbox{$<$}}}}
\def\grtsim{\mathrel{\hbox{\rlap{\hbox{\lower4pt\hbox{$\sim$}}}\hbox{$>$}}}}

\abstract{
The positions of more than 1300 nova eruptions in M31 catalogued
through the end of calendar year 2025 have been compared
in order to identify recurrent nova candidates.
The work extends the study of Shafter et al. (2015) who
identified a total of 12 recurrent novae with high confidence
(plus four possible recurrent novae) from
an analysis of 964 M31 novae observed prior to 2014.
During the past 12 years
an additional seven recurrent novae have been discovered in M31.
In addition, we have confirmed that one of the possible recurrent novae
is in fact recurrent (M31N 1990-10a),
while another was shown to be a foreground
dwarf nova (M31N 1966-08a). At present, there are
a total of 79 nova eruptions associated with
20 known recurrent novae in M31, with four additional eruptions from two
candidates remaining unconfirmed.
A comparison of the spatial distribution of the recurrent novae with
that for all novae shows no significant difference
between the two. In addition, we find no significant
difference between the light curve properties (peak luminosities
and rates of decline) between the M31 and Galactic recurrent nova populations.
However, the recurrence time distributions appear different,
with half of the M31 recurrent novae having recurrence times shorter than
U~Sco, the Galactic recurrent nova with the shortest known
recurrence time, $T_\mathrm{rec}=10.3$~yr. As expected,
recurrent novae are found to be both fainter and faster
than novae generally,
being mostly found in the lower left quadrant of the MMRD plane.
}

\FullConference{The Golden Age of Cataclysmic Variables and Related Objects - VII (GOLDEN2025)\\
1-6 September, 2025\\
Mondello, Palermo, Italy\\}

\begin{document}

\maketitle

\section{Introduction}

Classical novae can achieve absolute
magnitudes $M_V\lessim -10$ at the peak of their eruptions
making them among the most luminous
transient sources in the sky. The progenitors of novae are all
close binary systems consisting of a white dwarf (WD) primary
accreting material from a late-type secondary star.
The secondary,
which can be either on the main sequence or evolved, fills its
Roche lobe and transfers the material via an accretion
disk to the WD primary. As material builds up on the surface
of the WD, the temperature and density at the base of the accreted layer
eventually becomes sufficiently high to initiate
H burning. If the accretion rate is sufficiently low
($\dot M \lessim 10^{-7}~M_\odot$~yr$^{-1}$)
the material can become degenerate prior to H ignition, resulting in
a thermonuclear runaway (TNR) that powers the nova eruption.

Following a nova eruption, and the explosive ejection of all or
part of the accreted material\footnote{In some instances
more than the total of mass accreted may be ejected, resulting
in a secular decrease in the WD mass.},
the accretion cycle repeats, with the next eruption occurring
after a time interval, $T_\mathrm{rec}$, that depends on the
properties of the progenitor binary.
Recurrence times vary primarily as a function of WD mass and accretion rate,
and can vary from of order a year to upwards of $10^5$~yr, or longer.
Although all novae are recurrent on some timescale,
systems with $T_\mathrm{rec} \lessim 100$~yr
that have been observed to have more than one eruption
are referred to as "Recurrent Novae" (RNe).

The nearby Andromeda galaxy, M31, at a distance of $\sim760$~kpc
\citep{li2021}, offers an excellent opportunity to
study a large and equidistant sample of novae across a range
of stellar populations
\citep[e.g.,][and references therein]{Shafter2001,Shafter2011,Rector2022,Clark2024}.
As of 2025 a total of more than 1300 transient sources thought to be
novae have been discovered in
M31\footnote{\tt See https://www.mpe.mpg.de/$\sim$m31novae/opt/m31/M31\_table.html
and https://ashafter.sdsu.edu/m31\_nova\_table.html},
going back more than a century
to the pioneering surveys by Hubble and collaborators
\citep{Ritchey1917,Hubble1929}. A decade ago, Shafter et al. \citep{Shafter2015}
did an exhaustive search of the historical record of nova
eruptions to identify and
study the RN population in M31.
In this paper, we update the list of known M31 RNe, and
review our current understanding of the RNe populations
in M31 and the Galaxy.

\begin{figure}
\centering
\includegraphics[width=0.8\textwidth]{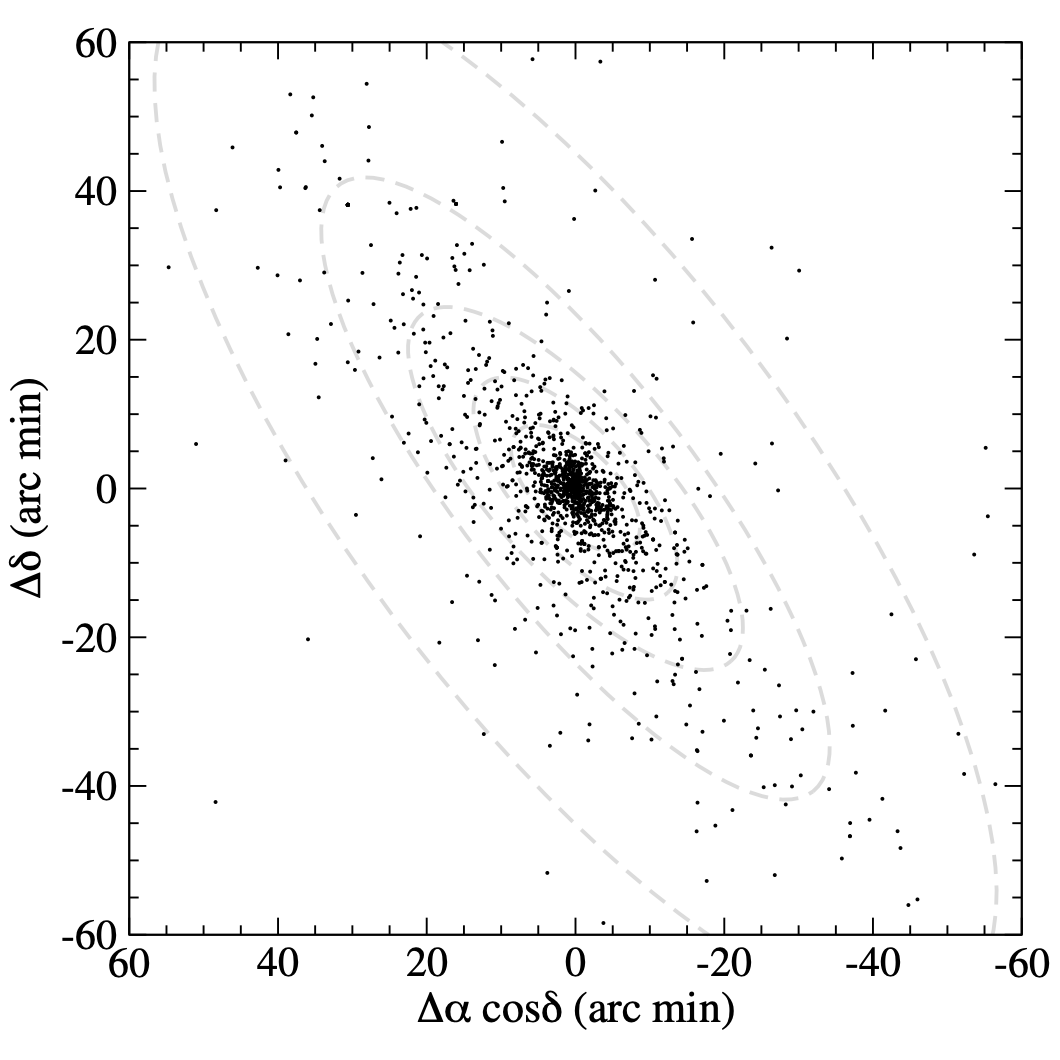}
\caption{The apparent distribution of 1334 nova candidates
observed in M31 between 1909 September and 2025 June. The
dashed grey lines show representative $R$-band isophotes.
Over time, the central (bulge) region of M31 has been surveyed
more frequently than the outer region of the galaxy. As a result,
the distribution does not accurately reflect the true spatial
distribution of M31 novae.
}
\label{fig1}
\end{figure}

\section{The M31 Nova Population}

Figure~\ref{fig1} shows the observed spatial distribution of 1334
M31 novae discovered between 1909 September, when the first
nova was discovered by Ritchey \citep{Ritchey1917}, and 2025 June.
Occasionally, long-period variables (LPVs or Mira variables)
have been found masquerading as novae. Thus,
a fraction of these objects should technically be regarded as
nova candidates if they have not been spectroscopically confirmed.
That said, in recent years spectroscopic follow-up has become common,
confirming that the vast majority of these luminous M31 transients
are in fact novae.

The nova discoveries over the years have come from
a wide range of surveys and observations with highly non-uniform
spatial coverage.
The early photographic surveys of M31 covered a large
fraction of the galaxy, but in contrast,
early CCD surveys, with their limited areal
coverage, focussed primarily on the bulge of M31
\citep[e.g.,][]{Ciardullo1987,Shafter2001}.
In recent years, large-format, automated
CCD surveys have once again extended the spatial coverage
\citep[e.g., the Zwicky Transient Facility,][]{Bellm2019}.
Given the non-uniform coverage,
the spatial distribution shown in Figure~\ref{fig1} cannot be assumed
to precisely represent the true distribution of novae
across M31. That said, the impression that novae are strongly
concentrated in the galaxy's bulge region is accurate.
When individual surveys of M31 with uniform spatial
coverage are analyzed independently, they have consistently shown
that novae are strongly concentrated in the central bulge region
of the galaxy \citep[e.g.,][]{Ciardullo1987,Shafter2001,Darnley2004,Rector2022}.

%\begin{figure}[!htb]
\begin{figure}
    \centering
    \includegraphics[width=0.6\textwidth]{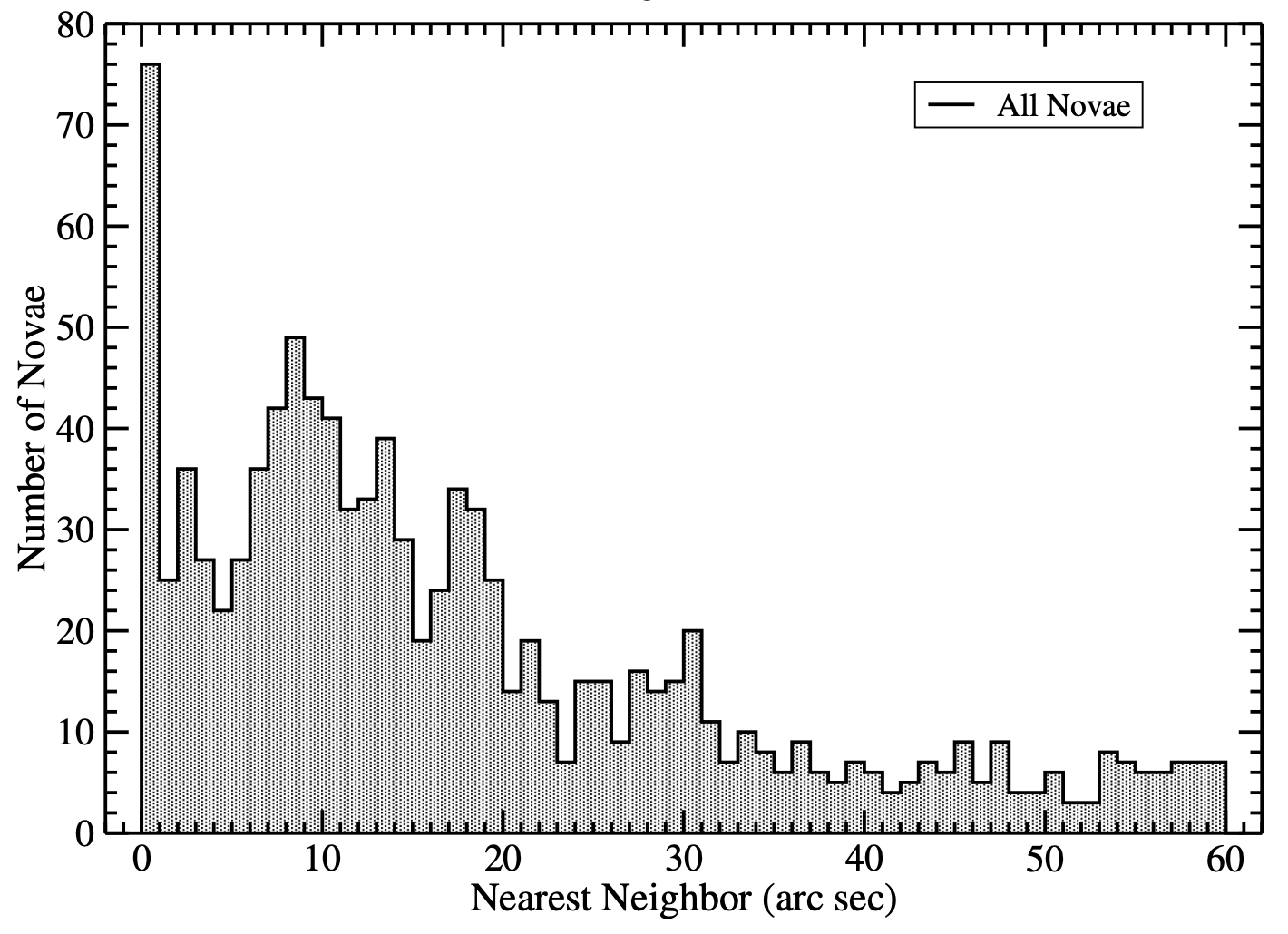}
    \includegraphics[width=0.6\textwidth]{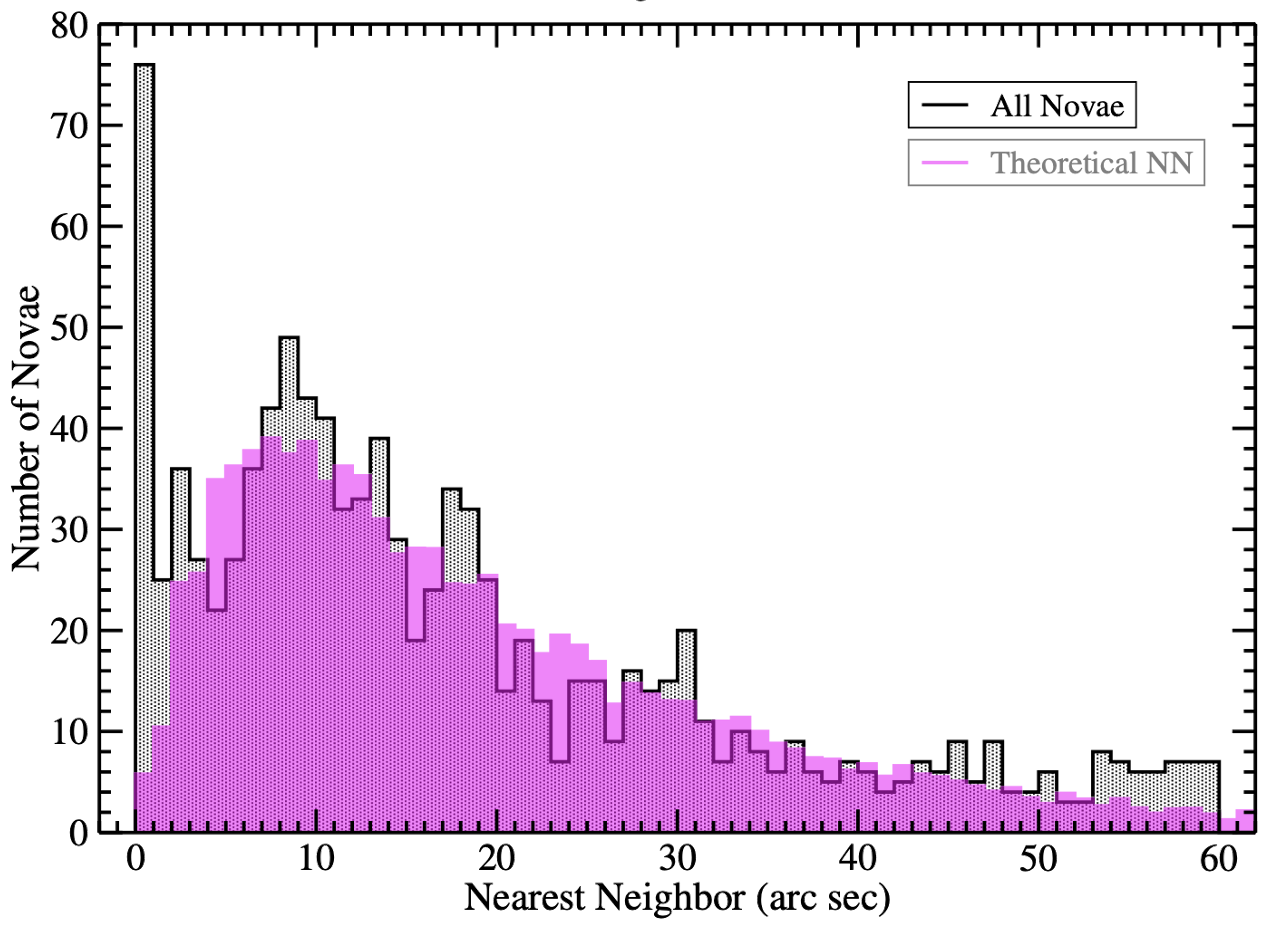}
    \includegraphics[width=0.6\textwidth]{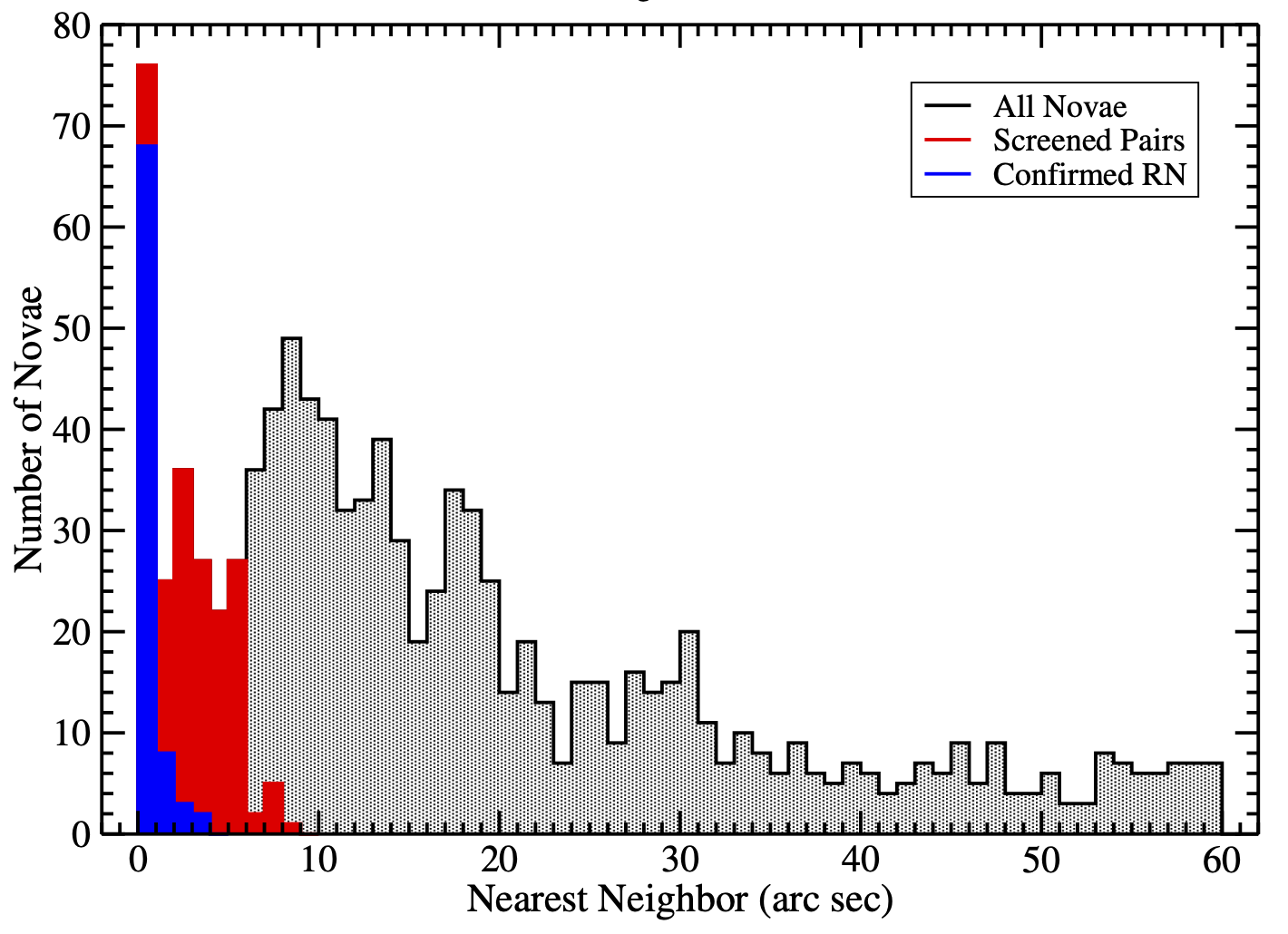}
    \caption{Top: The observed nearest-neighbor distribution
for all M31 nova candidates. Middle: The observed nearest-neighbor
distribution compared with a simulated nearest-neighbor distribution
containing no RN. Bottom: The nearest-neighbor distribution
with the screened novae (red) and the 22 RNe (blue) shown separately.}
    \label{fig2}
\end{figure}

\section{The Recurrent Nova Population of M31}

The wealth of nova data available in the historical record that
goes back more than a century offers an unparalleled opportunity
to study the RN population of M31.
In principle, the search for RNe is straightforward: the
published coordinates are cross-correlated to search for positional
coincidences that would identify multiple (two or more) outbursts arising
from the same progenitor binary. Of course, given
the inherent uncertainty in the reported positions, it
is much more difficult to distinguish between a true
spatial coincidence, and chance a positional coincidence
of two or more unrelated novae. The only way to establish that
a system is in fact recurrent is to acquire
images of the candidate RN eruptions and then carefully register
them to test for true spatial coincidence. In practice,
this requires digging up archival CCD images spanning the past
$\sim40$~yr, along with photographic plates going all the way
back to the early 20th century.

With this goal in mind, a decade ago Shafter et al. \citep{Shafter2015}
undertook the first systematic effort to identify the RN
population of M31. The positions for a total of 964 nova outbursts
in (up through and including calendar year 2013) were studied
to look for spatial coincidences that would suggest the possibility
that two or more eruptions arose from the same progenitor system.
At that time a total of 32 eruptions were ultimately determined
to have arisen from a total of 12 different RN progenitor systems,
with another 8 eruptions associated with an additional four possible RN
progenitors.

In the present paper, we extend the work of Shafter et al. \citep{Shafter2015}
to include the nearly 400 additional novae that have been recorded in
M31 over the past dozen years. As a first step, following
Shafter et al., it is instructive to consider
a nearest-neighbor (NN) distribution (see Figure~\ref{fig2})
determined for the entire M31 nova population. The distribution is
formed by going through the set of novae one-by-one and registering the
angular distance to its nearest neighbor nova. The resulting angular
separations were then placed into bins $1''$ wide for separations between
$s=1''$ and $s=60''$, with the number of novae
falling in each bin plotted
as a function of angular separation to form the distribution.

The upper panel of Figure~\ref{fig2} shows the updated NN distribution for
novae in M31.
The distribution is characterized by a broad, asymmetric peak centered at
a separation near $s=10''$ along with a spike in the $s=1''$ bin where we
expect many RNe will lie. It is interesting to compare the
observed NN distribution with
a simulated NN distribution computed from a sample of artificial novae
whose spatial distribution follows that of the overall M31 nova population,
but does not contain any RNe (i.e., there are no two simulated novae
at the same location).
The middle panel of Figure~\ref{fig2} shows such a distribution
compared with the actual distribution of M31 novae.
Both distributions show the broad asymmetric peak near $s=10''$,
that is characteristic of the overall
nova spatial distribution in M31, but only the observed NN distribution
displays the clear excess of novae with separations, $s\lessim3''$.

\begin{figure}
\centering
\includegraphics[width=7cm]{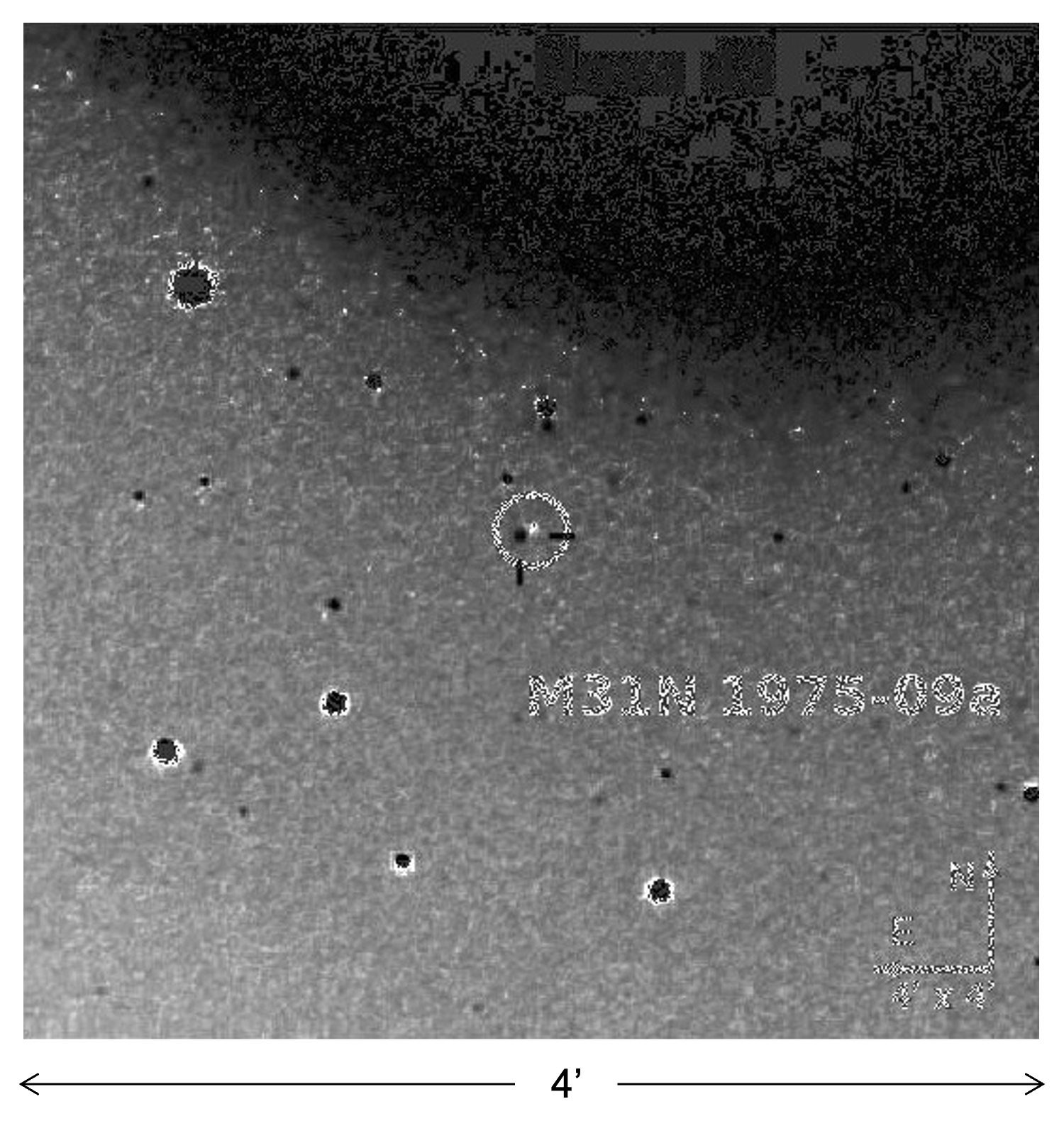}\qquad
\includegraphics[width=7cm]{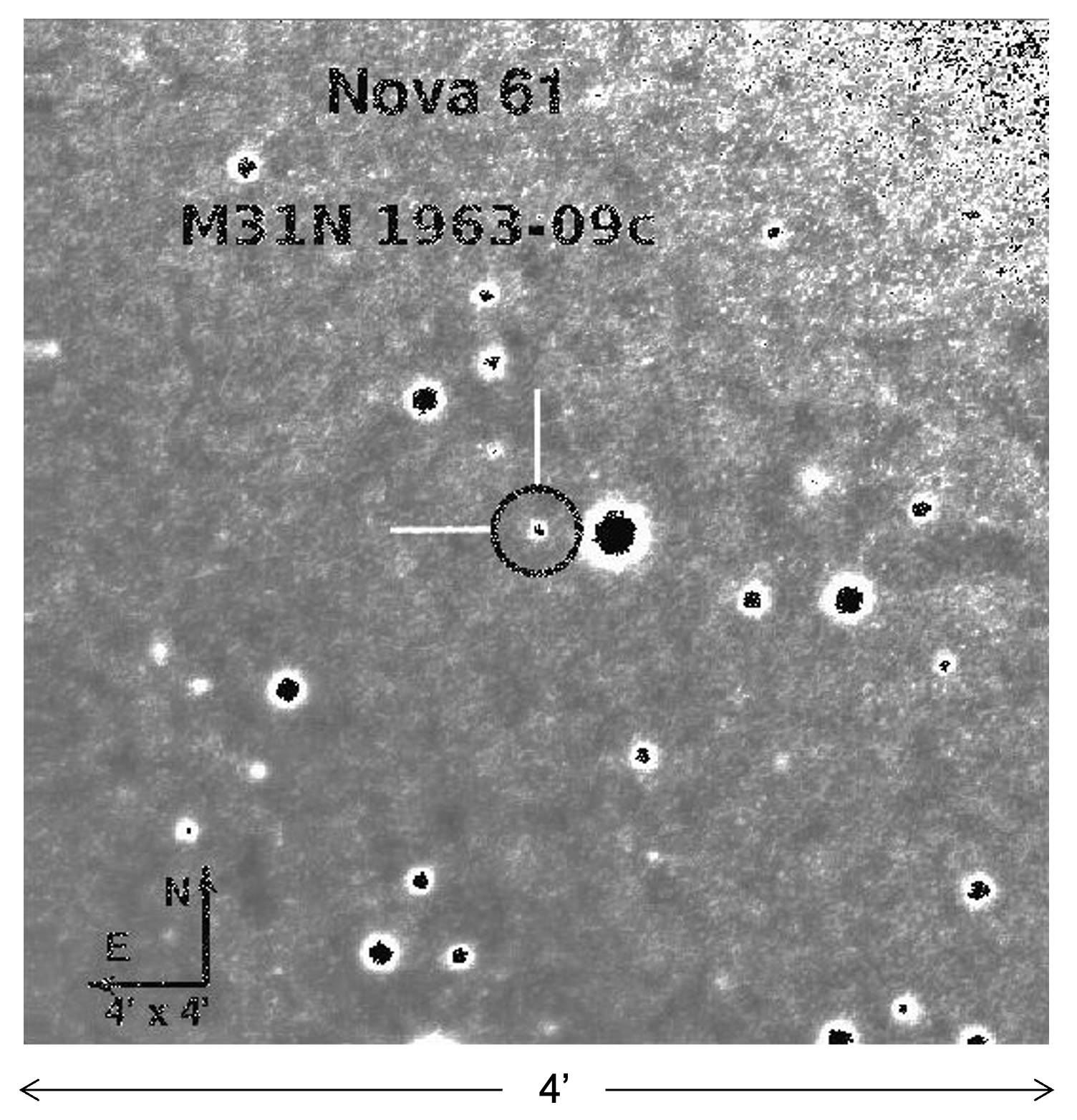}
\caption{Comparison images for two RN candidates from
Shafter et al. \citep{Shafter2015}. Left panel: M31N 1975-09a
compared with that of a possible recurrence M31N 1999-01a.
A careful comparison of the images show that these two novae
arise from distinct progenitors. Right panel: The
RN M31N 1963-09a compared with one of its six known
recurrences, M31N 2010-10e. 
}
\label{fig3}
\end{figure}

\subsection{The initial screening of potential RN candidates}

Despite the expectation that most novae with a nearest neighbor
a distance $s\lessim2''$ away will likely be recurrent,
it is possible (even likely near the center of M31),
that some fraction will turn out to be chance positional
coincidences of unrelated novae. On the other hand, given the
large positional uncertainties reported for many novae, especially
ones discovered in photographic surveys prior to 1980,
it is possible that some novae with separations as
considerably larger than $s=2''$ could upon closer inspection
be found to be spatially coincident and
associated with the same progenitor system.
Thus, in order to minimize the chance of missing a potential
RN, we routinely follow the coarse screening criteria 
used the Shafter et al. study to evaluate new potential RNe in M31.
If a new eruption is coincident to within $0.1'$ of
a previous eruption observed after 1980, it passes our screen
for further scrutiny. If the only previous eruption was one
detected with photographic plates, we relax the screen to $0.15'$.
The red distribution in the bottom panel of Figure~\ref{fig2}
illustrates the number of
novae screened in our most recent NN distribution.

\subsubsection{Probability of Chance Positional Coincidence}

%In an attempt to quantify the probability
%that two novae observed with a separation $s$
%are, in fact, outbursts from two unrelated progenitor systems,
Following Shafter et al. \citep{Shafter2015}
we can roughly estimate the probability, $P_\mathrm{C}$, of a
chance positional near-coincidence of two unrelated novae.
Intuitively, $P_\mathrm{C}$,
will depend on the observed nova surface density
(assumed to follow the background
$R$-band light of the galaxy)
in the vicinity of the nova.
We begin by summing the number of novae in
an elliptical annulus centered on the nucleus of
M31 that contains the position of the nova.
The width of the annulus is
chosen to be 1$'$, centered on the
nova's position. The nova surface density is then simply
given by the number of novae observed in the annulus, $N$,
divided by the area of the annulus, $A$.
Since we do not know {\it a priori\/}
which novae will be RN candidates,
the computation of $P_\mathrm{C}$ must consider the probability
that any nova in the annulus lies within a distance $s$ of any other nova
in the annulus. In practice, $P_\mathrm{C}$, for a given nova pair
is simply a function of the observed
separation, $s$, and the surface density of novae at the position in question.
We have:

\begin{equation}
P_\mathrm{C} \simeq 1 - \prod_{k=1}^{N-1}~(1 - kx),
\end{equation}

\noindent
where $x=\pi s^2/A$ is the fraction of the annulus area $A$ taken up
by each of the $N$ novae in the annulus.
Not surprisingly, the probability of a chance positional near-coincidence
is particularly high close to the center of M31
where the nova density is the highest (see Fig.~\ref{fig1})\footnote{Note that Equation~(1) slightly
overestimates $P_\mathrm{C}$, particularly in the crowded inner bulge region,
because it ignores both overlaps in exclusion zones and boundary effects.}.
While a low $P_\mathrm{C}$ is generally indicative of a RN,
a high $P_\mathrm{C}$ does not necessarily rule out a RN given the
uncertainties associated with many published nova coordinates.

Although computation of $P_\mathrm{C}$ can add confidence to
our assessment of a RN candidate passing our coarse screen,
original images of the eruptions must be found and carefully registered
in order to firmly establish whether the novae in question are in fact
spatially coincident.
Figure~\ref{fig3} illustrates two examples of this process
from the earlier Shafter et al. \citep{Shafter2015} study. 
The RN candidate shown in the left panel, M31N 1975-09a, is
compared with a possible recurrence M31N 1999-01a. The two novae
were found to be unrelated. The right panel of Figure~\ref{fig3}
shows the RN M31N 1963-09c compared with a recurrence observed
in 2010 (M31N 2010-10e). The novae are coincident to better than
an arc second, which at an isophotal (de-projected) distance $a=17.6'$ from
the center of M31, reduces the probability of a chance positional
coincidence of unrelated novae to $\lessim0.0002$. Since the
publication of the Shafter et al. study in early 2015, an additional
3 eruptions of this RN have now been observed (M31N 2015-10c, 2020-11b,
and 2025-11d). M31N 1963-09c has now been observed to erupt
more often than any other RN extant, with the exception of
M31N 2008-12a, the well-known RN with an unprecedented 1~yr recurrence time.

A total of 12 of the screened novae
studied by Shafter et al. \citep{Shafter2015}
were judged to be recurrent, with another four being possibly recurrent.
Of the latter four, one system (M31N 1990-10a)
has now displayed two additional outbursts,
confirming it to be a RN, while
another system (M31N 1966-08a) has been found to be a
foreground dwarf nova \citep{Shafter2017b}.
The other two systems, M31N 1953-09b and 1961-11a, remain likely,
but unconfirmed RNe.

\begin{deluxetable}{lcrccrcr}
\tabletypesize{\scriptsize}
\tablenum{1}
\thispagestyle{empty}
\tablewidth{0pt}
\tablecolumns{8}
\tablecaption{M31 Recurrent Novae (excluding M31N 2008-12a)\label{tab1}}
\tablehead{\colhead{Recurrent Nova} & \colhead{Recurrence} & \colhead{$\Delta t$ (yr)} & \colhead{$M_\mathrm{R}$(max)} & \colhead{$t_2$ (d)} & \colhead{$a$ ($'$)} & \colhead{$P_\mathrm{C}$} & \colhead{Type\tablenotemark{a}}
}
\startdata
M31N1919-09a & M31N1998-06a &78.7 &\dots         &\dots         &  10.50 &0.021600 &  \dots  \cr
M31N1923-12c & M31N2012-01b &88.1 &$-6.96$       &$12.6\pm0.9$  &  10.99 &0.005484 &  He/N   \cr
             & M31N2021-07a & 9.5 &\dots         &\dots         &  11.00 &0.003376 &  He/N   \cr
M31N1926-06a & M31N1962-11a &36.5 &\dots         &$\grtsim$5    &  17.43 &0.009649 &  \dots  \cr
M31N1926-07c & M31N1997-10f &71.3 &\dots         &\dots         &   1.68 &0.059698 &  \dots  \cr
             & M31N2008-08b &10.8 &\dots         &\dots         &   1.69 &0.012253 &  He/Nn  \cr
             & M31N2020-01b &11.4 &$-7.16$       &$11.4\pm2.5$  &   1.67 &0.103444 &  He/Nn: \cr
             & M31N2022-09a & 2.7 &$-6.86$       &$10.8\pm1.5$  &   1.68 &0.088261 &  \dots  \cr
M31N1945-09c & M31N1975-11a &27.0 &\dots         &\dots         &  29.19 &0.000170 &  \dots  \cr
M31N1953-09b: & M31N2004-08a &50.8 &\dots        &\dots         &   5.77 &0.234561 &  \dots  \cr
M31N1960-12a & M31N2013-05b &52.4 &$-6.76$       &$4.7\pm0.2$   &   4.40 &0.149255 &  He/N   \cr
             & M31N2019-07a & 6.2 &\dots         &\dots         &   4.40 &0.061079 &  \dots  \cr
M31N1961-11a: & M31N2005-06c &43.6 &\dots        &\dots         &   3.03 &0.063232 &  \dots  \cr
M31N1963-09c & M31N1968-09a & 5.0 &\dots         &$\grtsim$4    &  17.60 &0.003105 &  \dots  \cr
             & M31N2001-07b &32.8 &\dots         &5:            &  17.60 &0.000643 &  \dots  \cr
             & M31N2010-10e & 9.3 &$-6.51$       &$4.3\pm0.5$   &  17.60 &0.000643 &  He/N   \cr
             & M31N2015-10c & 5.0 &$-6.71$       &$3.6\pm0.5$   &  17.59 &0.000097 &  He/N   \cr
             & M31N2020-11b & 5.1 &\dots         &\dots         &  17.60 &0.000584 &  He/N   \cr
             & M31N2025-11d & 5.0 &\dots         &\dots         &  \dots &\dots    & \dots   \cr
M31N1966-09e & M31N2007-08d &40.9 &$-6.06$       &$13.0\pm1.5$  &  59.56 &0.000083 &  Fe II: \cr
M31N1982-08b & M31N1996-08c &15.0 &\dots         &\dots         &  60.83 &0.000129 &  \dots  \cr
M31N1984-07a & M31N2004-11f &20.3 &\dots         &$28.4$        &   0.57 &0.057987 &  \dots  \cr
             & M31N2012-09a & 7.8 &$-8.26$       &$9.7\pm1.4$   &   0.57 &0.008155 &  FeIIb  \cr
             & M31N2022-11b &10.2 &$-8.43$       &$9.8\pm1.6$   &   0.57 &0.145043 &  \dots  \cr
M31N1990-10a & M31N2007-07a &16.7 &\dots         &$\lessim$10   &   4.28 &0.077728 &  \dots  \cr
             & M31N2016-07e & 9.0 &$-7.00$       &$11.1\pm1.7$  &   4.27 &0.009891 &  FeIIb  \cr
             & M31N2022-03b & 5.7 &\dots         &\dots         &   4.29 &0.103048 &  \dots  \cr
M31N1997-11k & M31N2001-12b & 4.2 &$-5.86$       &$\grtsim$100  &  \dots &\dots    &  \dots  \cr
             & M31N2009-11b & 7.9 &$-5.96$       &$110\pm29$    &  10.57 &0.000205 &  Fe II: \cr
             & M31N2021-08a &11.8 &$-5.86$       &\dots         &  10.47 &0.080037 &  \dots  \cr
M31N2001-11a & M31N2025-02b &23.2 &$-$7.5:       &8:            &  27.90 &0.000442 &  \dots  \cr
M31N2005-10a & M31N2022-08a &16.8 &\dots         &\dots         &  29.43 &0.000870 &  \dots  \cr
M31N2006-11c & M31N2015-02b & 8.2 &$-8.02$       &$1.8\pm0.5$   &  19.82 &0.002533 &  He/N   \cr
M31N2007-10b & M31N2017-12a &10.2 &$-6.21$       &$2.7\pm0.6$   &  12.90 &0.000736 &  He/N   \cr
M31N2007-11f & M31N2016-12e & 9.1 &$-7.22$       &$10.2\pm2.6$  &  18.90 &0.002937 &  He/Nn  \cr
M31N2013-10c & M31N2023-11f &10.1 &$-8.66$       &$  5.5\pm1.7 $&   6.38 &0.534439 &  \dots  \cr
M31N2017-01e & M31N2012-01c &$-5.1$&\dots        &\dots         &  \dots &\dots    &  He/N   \cr
             & M31N2014-06c &$-2.6$&$-6.66$      &\dots         &  \dots &\dots    &  \dots  \cr
             & M31N2019-09d & 2.7 &$-6.86$       &$\sim6$       &  54.19 &0.000006 &  \dots  \cr
             & M31N2022-03d & 2.5 &$-6.66$       &\dots         &  54.20 &0.000028 &  \dots  \cr
             & M31N2024-08c & 2.4 &$-6.76$       &$5.2\pm1.6$   &  \dots &0.000004 &  He/N   \cr
\enddata
\tablenotetext{a}{Spectroscopic type of the nova and/or its recurrence.}
\end{deluxetable}

\begin{deluxetable}{lcrccrcr}
\tabletypesize{\scriptsize}
\tablenum{2}
\thispagestyle{empty}
\tablewidth{0pt}
\tablecolumns{8}
\tablecaption{M31N 2008-12a\label{tab2}}
\tablehead{\colhead{Recurrent Nova} & \colhead{Recurrence} & \colhead{$\Delta t$ (yr)} & \colhead{$M_\mathrm{R}$(max)} & \colhead{$t_2$ (d)} & \colhead{$a$ ($'$)} & \colhead{$P_\mathrm{rec}$} & \colhead{Type\tablenotemark{a}}
}
\startdata
M31N2008-12a & M31N1992-02a &$-15.9$&\dots       &\dots         &  48.92 &0.048289 &  \dots  \cr 
             & M31N1993-01b &$-15.0$&\dots       &\dots         &  48.92 &0.048289 &  \dots  \cr 
             & M31N2001-09a &$-8.3$ &\dots       &\dots         &  48.92 &0.001377 &  \dots  \cr
             & M31N2009-12b &  1    &\dots       &\dots         &  48.92 &  \dots  \cr
             & M31N2010-11b & 0.9   &\dots       &\dots         &  48.92 &0.000216 &  \dots  \cr
             & M31N2011-10e & 0.9   &$-6.36$     &\dots         &  48.92 &0.000385 &  \dots  \cr
             & M31N2012-10a &  1    &$-6.16$     &\dots         &  48.92 &0.002412 &  He/N   \cr
             & M31N2013-11f & 1.1   &$-6.26$     &\dots         &  48.92 &0.004480 &  He/N   \cr
             & M31N2014-10c & 0.9   &$-6.32$     &$3.42\pm0.40$ &  48.92 &0.000216 &  He/N   \cr
             & M31N2015-08d & 0.8   &$-6.24$     &$1.96\pm0.12$ &  48.92 &0.000216 &  He/N   \cr
             & M31N2016-12f & 1.3   &$-6.80$     &$2.46\pm0.18$ &  48.93 &0.004209 &  He/N   \cr
             & M31N2017-12e &  1    &$-6.38$     &$4.44\pm0.38$ &  48.93 &0.004209 &  He/N   \cr
             & M31N2018-11a & 0.9   &$-6.21$     &$2.49\pm0.34$ &  48.93 &0.004209 &  He/N   \cr
             & M31N2019-11b &  1    &$-6.54$     &$2.33\pm0.63$ &  48.93 &0.004209 &  He/N   \cr
             & M31N2020-10g & 0.9   &$-6.26$     &\dots         &  48.92 &0.000041 &  He/N   \cr
             & M31N2021-11b & 1.1   &$-6.31$     &$2.40\pm0.07$ &  48.92 &0.001954 &  He/N   \cr
             & M31N2022-12a & 1.1   &$-6.30$     &\dots         &  48.93 &0.005512 &  He/N   \cr
             & M31N2023-12b & 1.0   &$-6.30$     &\dots         &  48.93 &0.005512 &  He/N   \cr
             & M31N2024-12a & 1.0   &\dots       &\dots         &  48.93 &0.005512 &  He/N   \cr
             & M31N2025-11e & 1.0   &\dots       &\dots         &  \dots &\dots    &  He/N   \cr
\enddata
\tablenotetext{a}{Spectroscopic type of the nova and/or its recurrence.}
\end{deluxetable}

\section{Properties of the M31 Recurrent Nova Population}

Continuing observations over the past 12 years have revealed seven new
RNe \citep{Hornoch2015,Sin2017,Shafter2018,Hornoch2022,Shafter2022a,Shafter2022b,Shafter2022c,Shafter2024a,Shafter2024b,Shafter2025},
bringing the total number in M31 to 22
systems (with two remaining uncertain).
%For the purposes of the subsequent discussion
%we will assume these two systems are in fact RNe, yielding a total
%of 22 RNe thus far identified in M31.
The properties of the individual RNe are summarized in Table~\ref{tab1}.
Many of the RNe have been observed to erupt multiple times, but
it is clear from looking at the variation in the
inter-eruption interval, $\Delta t$,
that many eruptions have been missed. Thus, for many of the RNe,
$\Delta t$ represents an upper limit to the true recurrence time.
Given the large number of recurrences observed
for M31N~1008-12a, we have summarized its properties separately
in Table~\ref{tab2}.
%represented by the blue region in the bottom panel of Figure~\ref{fig2},

\subsection{The Spatial Distribution}

\begin{figure}
\centering
\includegraphics[angle=0,scale=0.17]{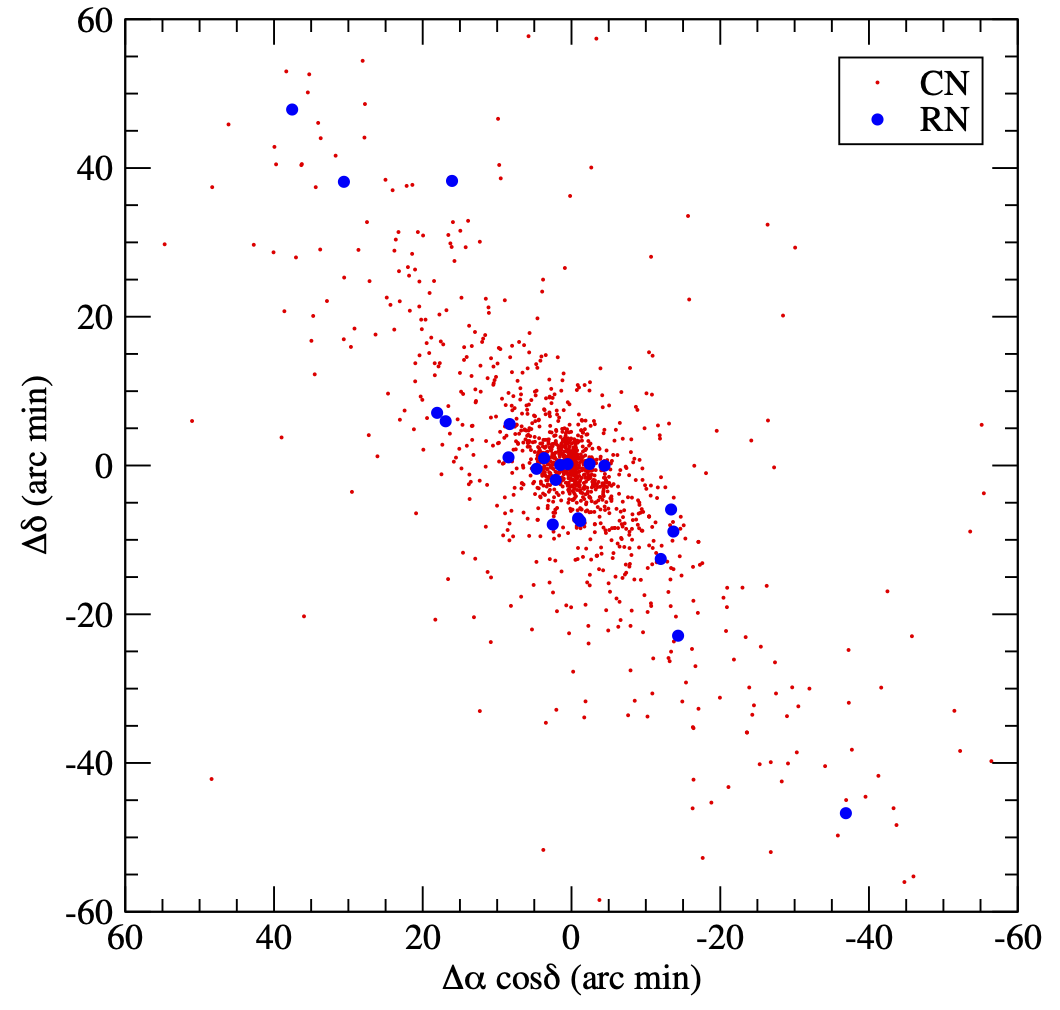}
\includegraphics[angle=0,scale=0.17]{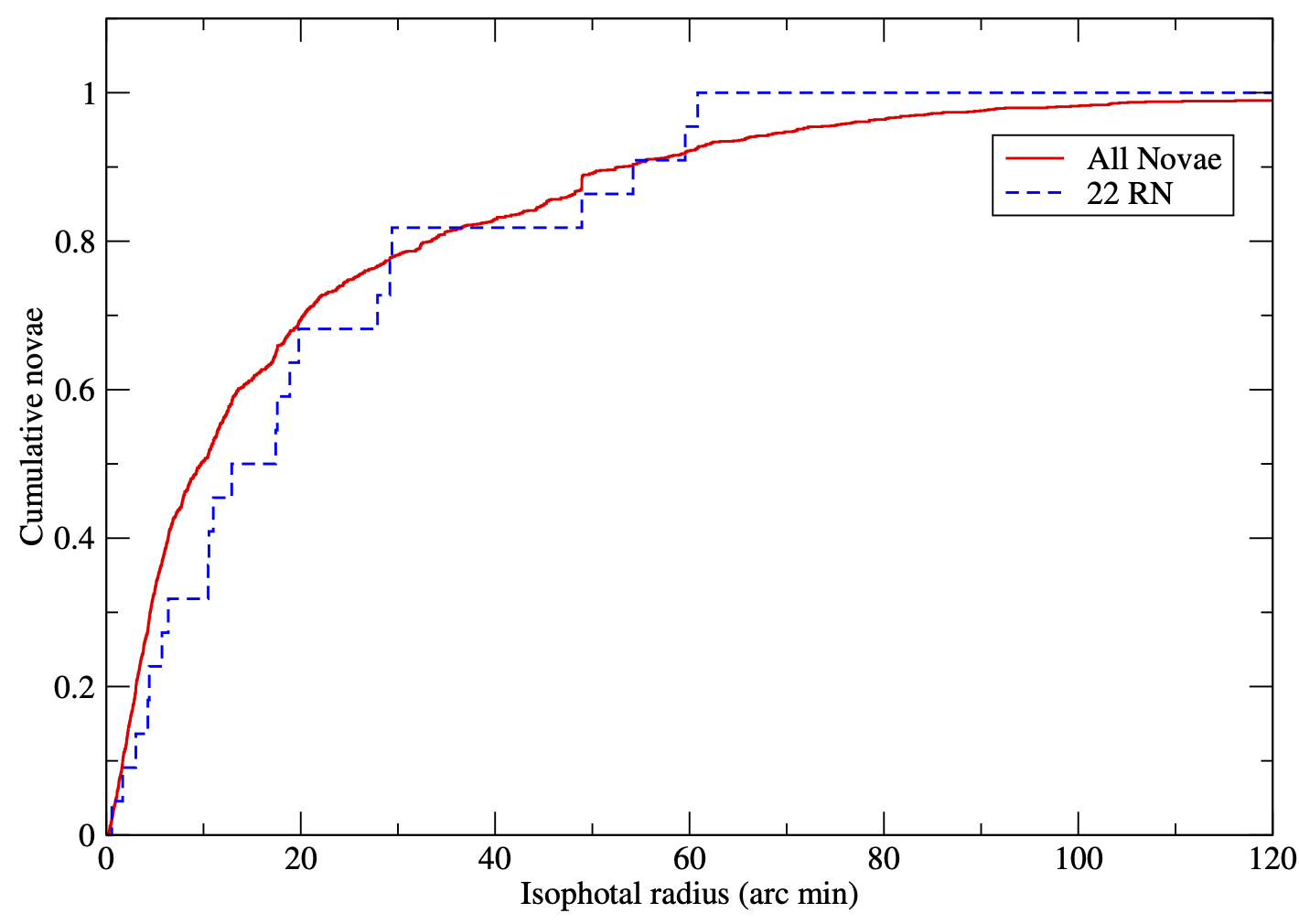}
\caption{Left: The spatial distribution of RN compared
with the overall nova distribution. Right: The cumulative
distribution of RN compared with all novae. A K-S test confirms
that there is no significant difference between the two distributions.
}
\label{fig4}
\end{figure}

Figure~\ref{fig4} shows the spatial distribution of the 22 RNe
compared with the overall classical nova population of M31 (CNe). In the
left panel, the spatial positions of the RNe and CNe
are plotted separately, while the
right panel shows the cumulative fractions of RNe and CNe
as functions of the $R$-band isophotal radius in M31. Although, it is
difficult to draw any firm conclusions about the relative
distributions of RNe and CNe from direct observation of their
spatial positions, the cumulative distributions show that
their spatial distributions are remarkably
similar with a K-S test (K-S $p=0.75$) showing no
significant difference between the two distributions.

\subsection{The Light Curve Properties}

\begin{figure}
\centering
\includegraphics[angle=0,scale=0.4]{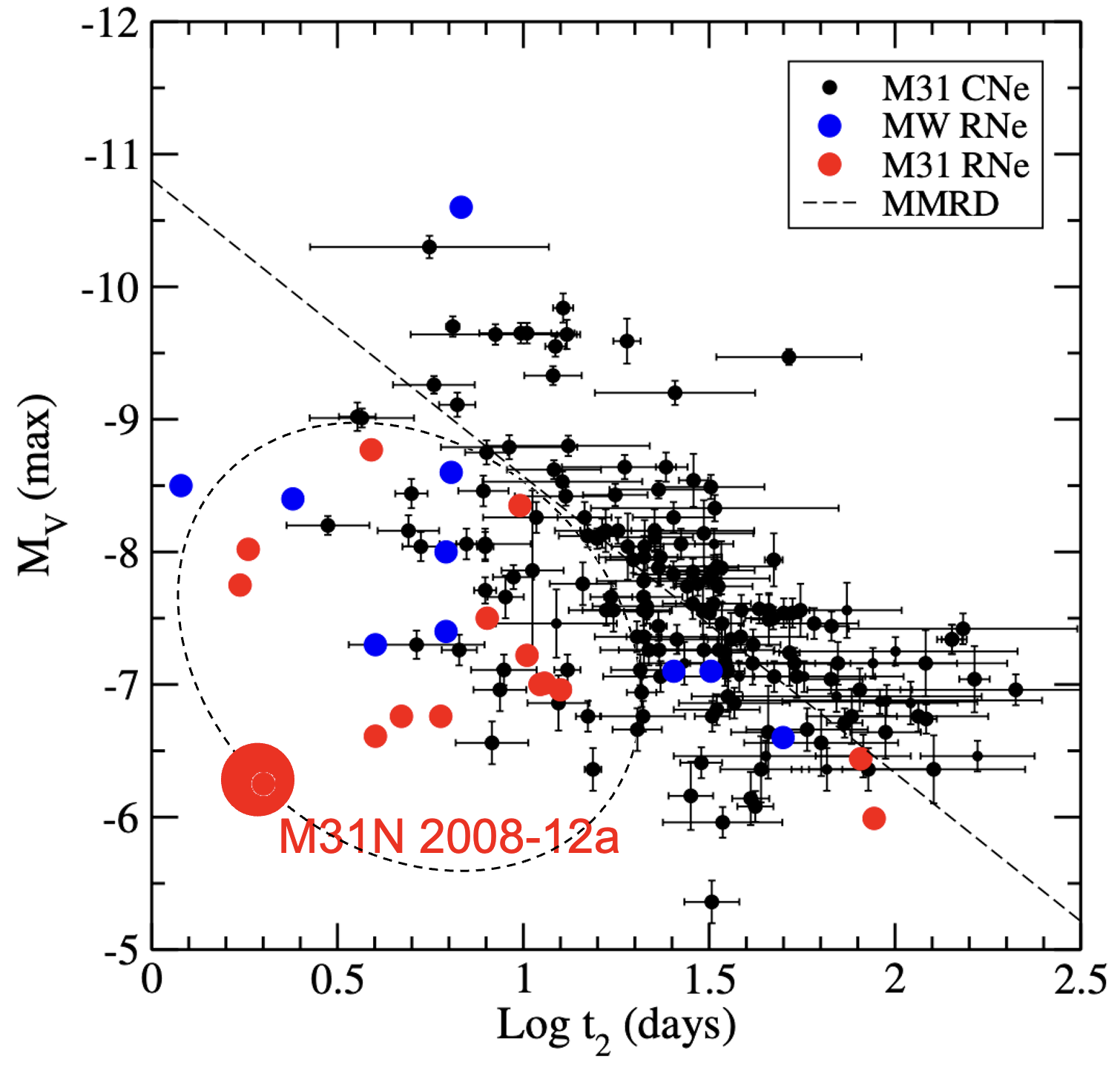}
\caption{The MMRD relation of the sample of novae from Clark et al. \citep{Clark2024}.
Galactic RNe are shown in blue and M31 novae in red. The RNe from
both galaxies are fainter and faster than novae
generally, and lie in the lower left quadrant of the MMRD diagram (shown
by the dashed ellipse). The two RNe near $\log t_2 = 1.9$ are both Fe~II class.
The one-year recurrence time RN,
M31N 2008-12a, with $M_V\sim-6.2$ and $t_2\sim2.1$~d, is highlighted.
}
\label{fig5}
\end{figure}

The relationship between the
observed light curve properties of novae, specifically
their peak luminosities and rates of decline, can best
be illustrated by considering the Maximum Magnitude
versus Rate of Decline (MMRD) relationship. The MMRD relation
dates back almost a century to when
McLaughlin \citep{McLaughlin1939,McLaughlin1942,McLaughlin1945}
first noticed that among classical novae, the most luminous
systems tended to fade the quickest.
The veracity of the MMRD relation has been called into question
in recent years, primarily because an increasing number of
low-luminosity novae have be discovered to fade quickly.
Such novae have been selected against in nova surveys lacking the
required depth and cadence necessary for their discovery.
Figure~\ref{fig5} shows
the latest MMRD relation for novae in M31, which has recently been
published by our group \citep{Clark2024}. The known RNe in M31 and
in the Galaxy are found overwhelmingly in the lower left quadrant of the
MMRD plot.

\begin{figure}
\centering
\includegraphics[angle=0,scale=0.4]{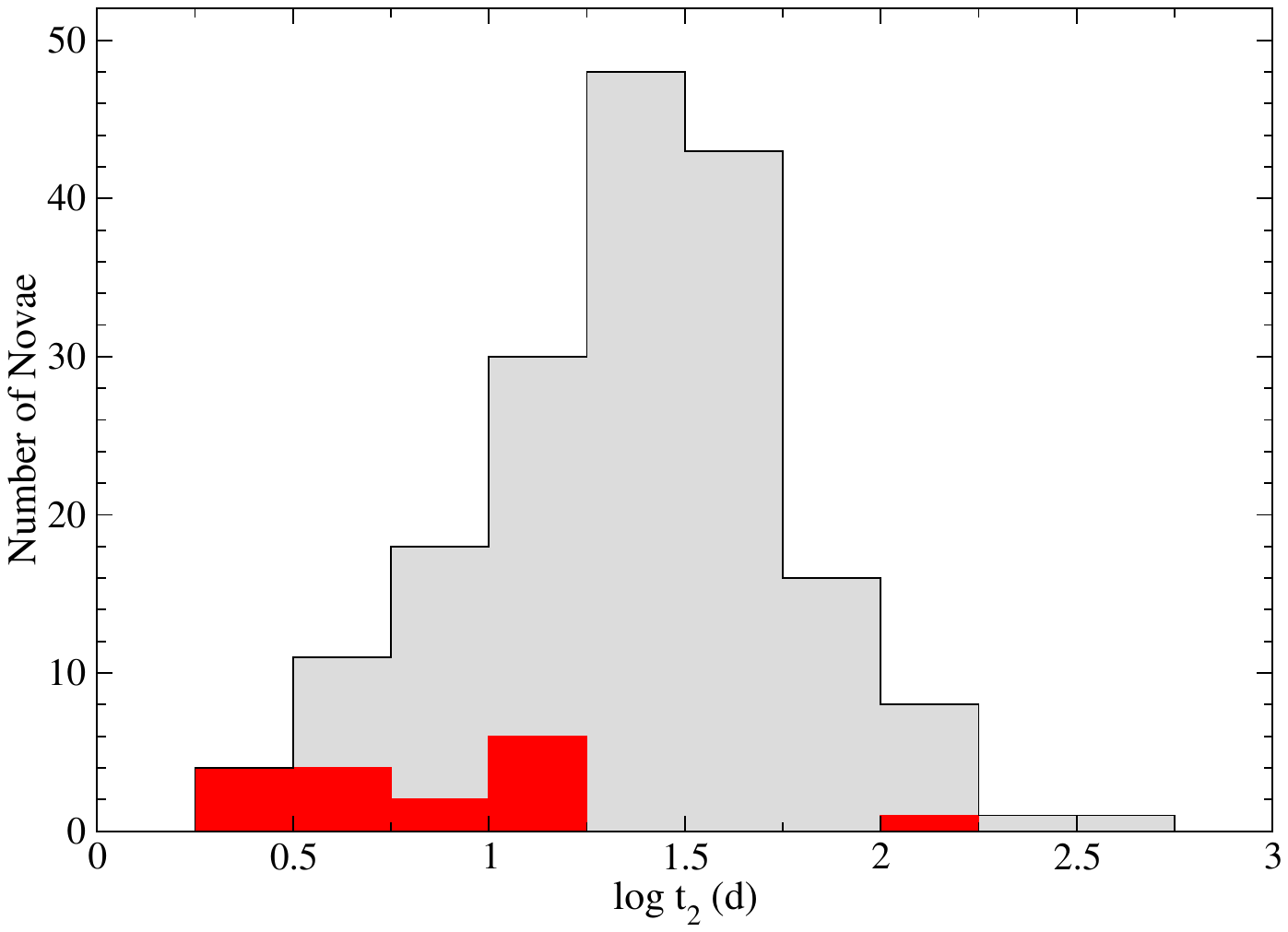}
\caption{The rates of decline ($\log t_2$) for the 17 RNe in the
Clark et al. \citep{Clark2024} sample of M31 novae compared with
novae generally. With one exception (M31N~2009-11b = 1997-11k),
the RNe are among the fastest declining novae. 
}
\label{fig6}
\end{figure}

Among novae, models show that systems with the shortest recurrence times
require massive WDs accreting at high rates \citep{Townsley2005,
Wolf2013,Kato2014}. The required ignition masses are relatively small
($M_\mathrm{ign}\lessim10^{-6}~M_\odot$~yr$^{-1}$ in the case of the
RNe with the shortest recurrence times),
and are built up
quickly with the hot accreted layer being only partially degenerate
at the onset of the TNR. The resulting eruption is relatively weak,
ejecting a relatively small amount of material.
The recurrence times $T_\mathrm{rec} (\equiv M_\mathrm{ign}/{\dot M}$)
can be as short as a year for accretion rates of
a few $\times~10^{-7}~M_\odot$~yr$^{-1}$
and WD masses approaching the Chandrasekhar limit.
Observationally, RNe are typically
characterized by relatively low peak luminosities ($M_V\grtsim-8$) with
rapid declines from maximum light ($t_2\lessim10$~d).
Thus, RNe are exactly the ``faint and fast" population that
depart strongly from the MMRD relation, as shown in Figure~\ref{fig5}.
The rapid rates of decline for RNe compared with novae generally, is
nicely illustrated in Figure~\ref{fig6}.

%The distribution of recurrence times for the known M31 RNe novae
%is shown in Figure~\ref{fig8}.

\begin{figure}
\centering
\includegraphics[width=7cm]{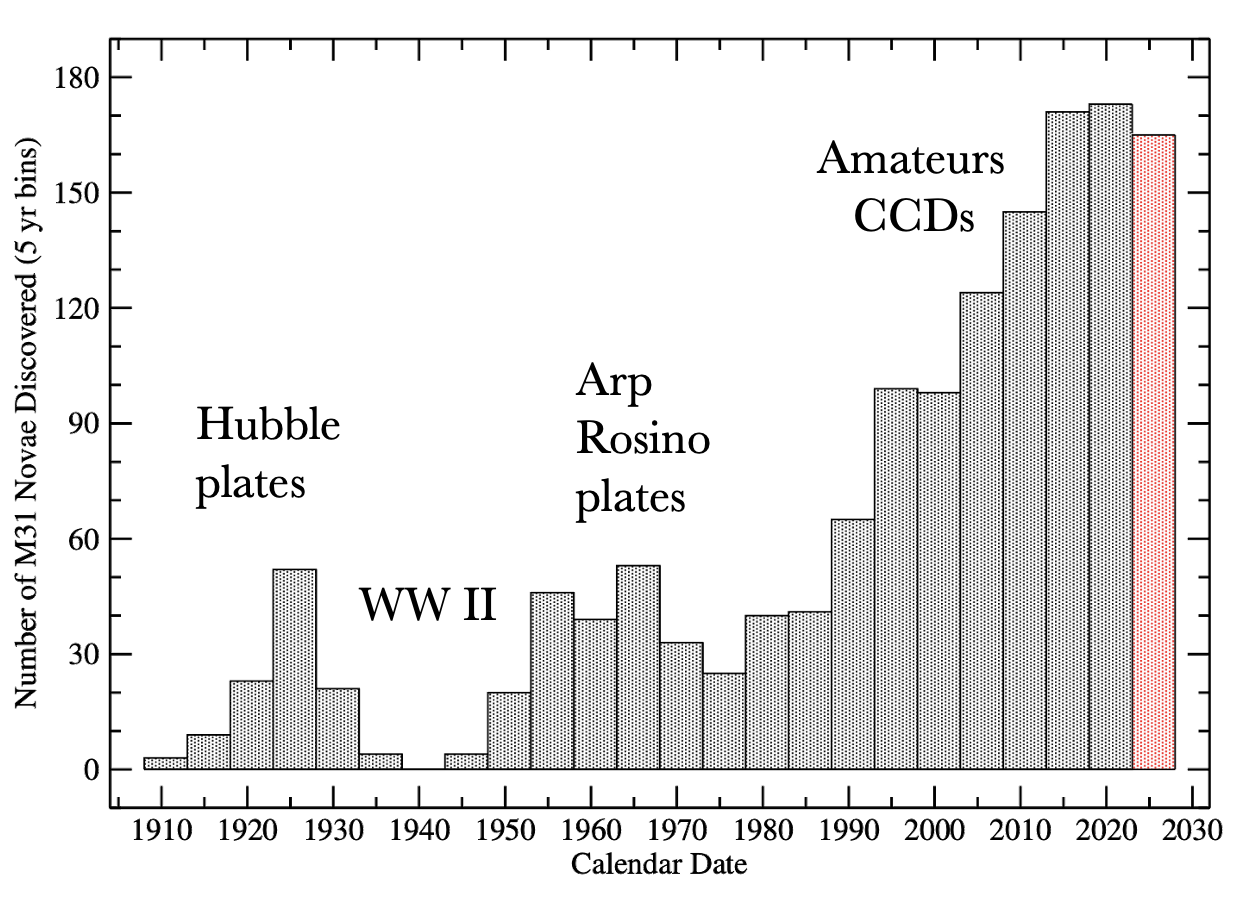}\qquad
\includegraphics[width=7cm]{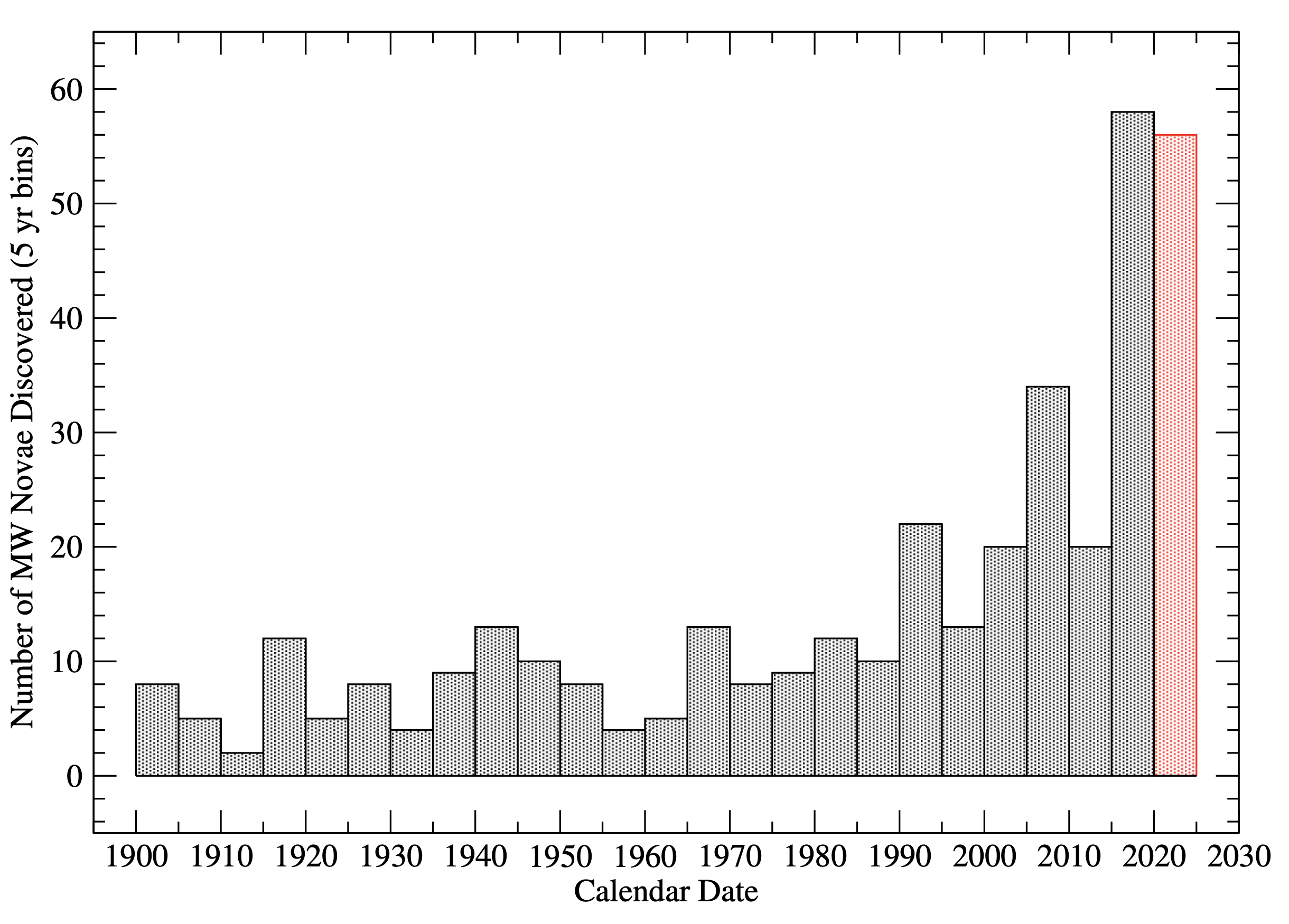}
\caption{The rate of discovery of M31 novae (left panel) compared with the
rate of discovery in the Galaxy (right panel). The red regions show the
expected discovery rate for the 2021-2025 period based on
an extrapolation of observations through June 2025.
The rate of discovery in the
Galaxy has been roughly one-third that for M31.
It appears that the discovery rate in M31 may be plateauing
at a level of $\sim35$~yr$^{-1}$, while that for the
Galaxy is still rising.
}
\label{fig7}
\end{figure}

\section{Comparison of the M31 and Galactic RN Populations}

Figure~\ref{fig7} shows the history of nova discoveries
going back a little over the past century
in M31 compared with the Milky Way.
Interestingly, the rate of discovery in M31
has outpaced that in the Galaxy by roughly a factor of 3,
uniformly across time. We expect that ratio to begin declining
rapidly as wide-field automated surveys (e.g., Pan-STARRS, ASAS-SN, ATLAS,
and now Rubin \citep{Chambers2016,Kochanek2017,Tonry2018,Ivezic2019}),
which will disproportionally
aid the discovery of Galactic novae, continue to be deployed.

In the case of M31, a total of
81 eruptions ($\sim6$\% of the total number observed)
have been associated with the 22 known RNe.
A total of only 10 Galactic RNe have thus far been identified out
of a total of $\sim400$ nova eruptions seen over this period,
with IM~Nor being
the last RN to be recognized nearly a quarter century ago \citep{Kato2002}. 
That said, the 10 known RNe have accounted for 43 of the estimated 400
eruptions ($\grtsim10$\%), roughly twice the percentage seen in M31. 

\begin{figure}
\centering
\includegraphics[angle=0,scale=0.34]{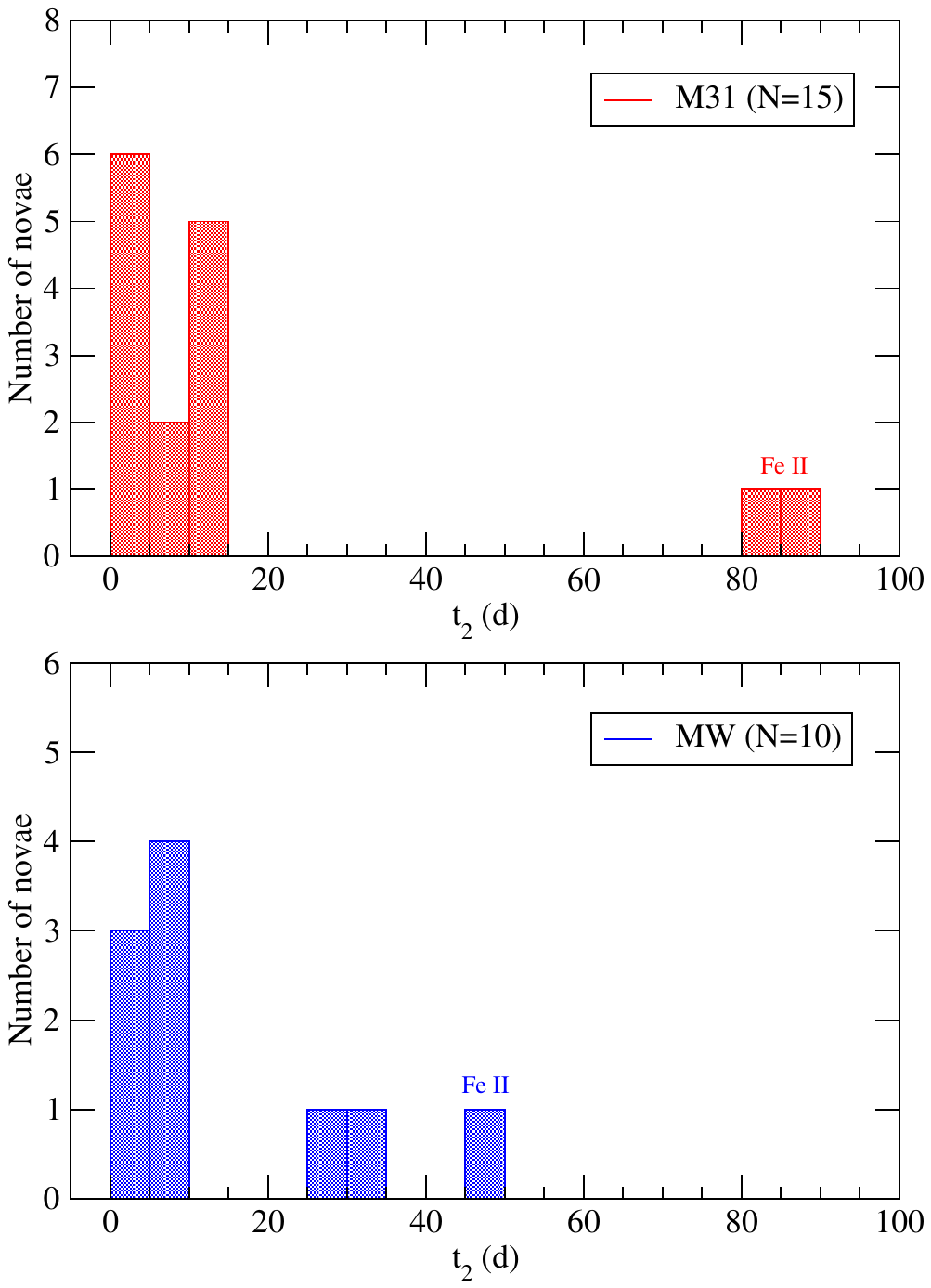}
\includegraphics[angle=0,scale=0.34]{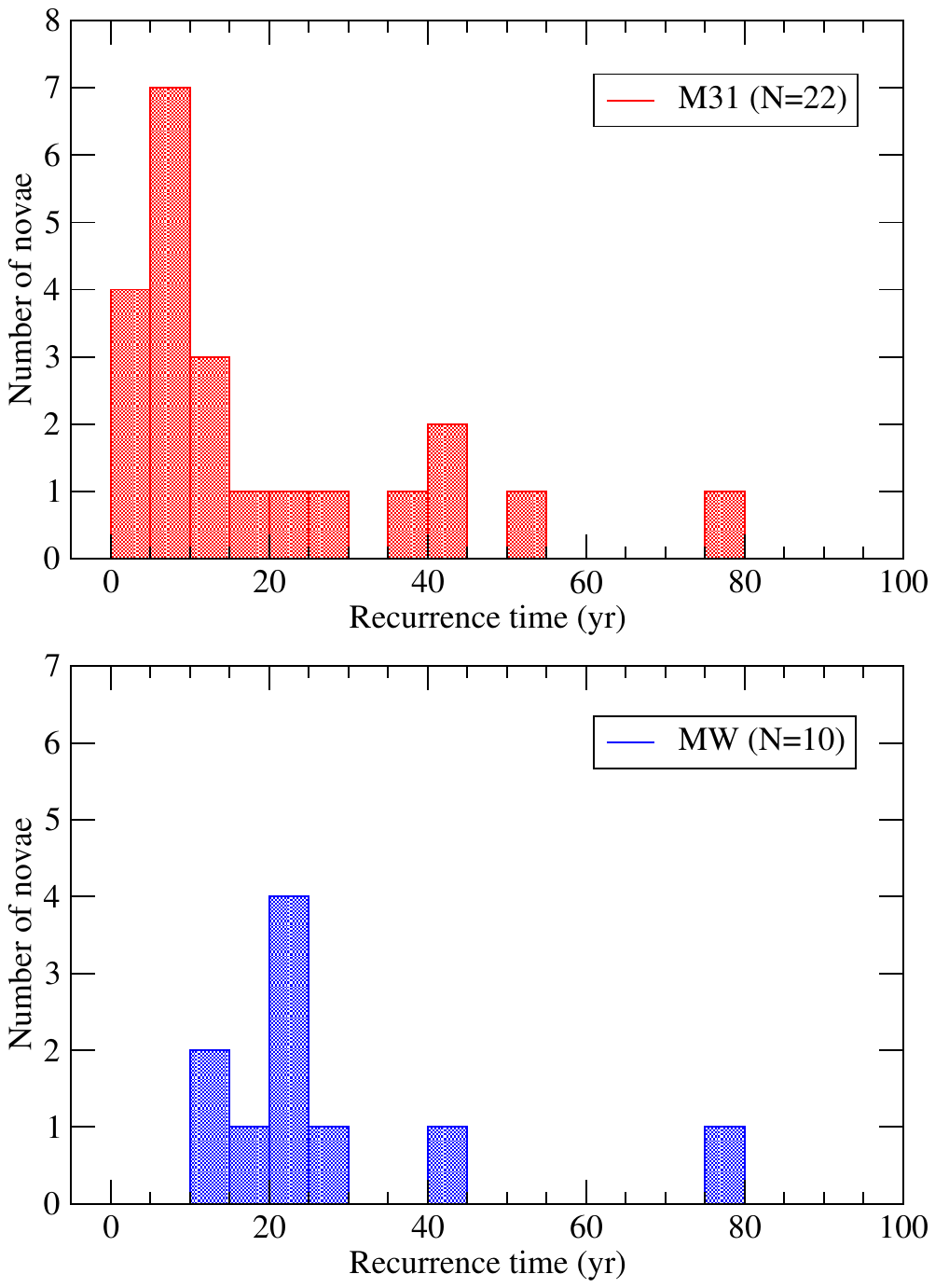}
\caption{
Left: The $t_2$ time distribution for the M31 RNe (top)
compared with the corresponding Galactic RNe distribution (bottom).
Right: The recurrence time distribution for
the 22 known M31 RN (top) compared with the distribution for the
10 known Galactic RN (bottom).
Half of the M31 RNe have recurrence times below
10~yr, while none of the Galactic novae do.
}
\label{fig8}
\end{figure}

\begin{figure}
\centering
\includegraphics[angle=0,scale=0.26]{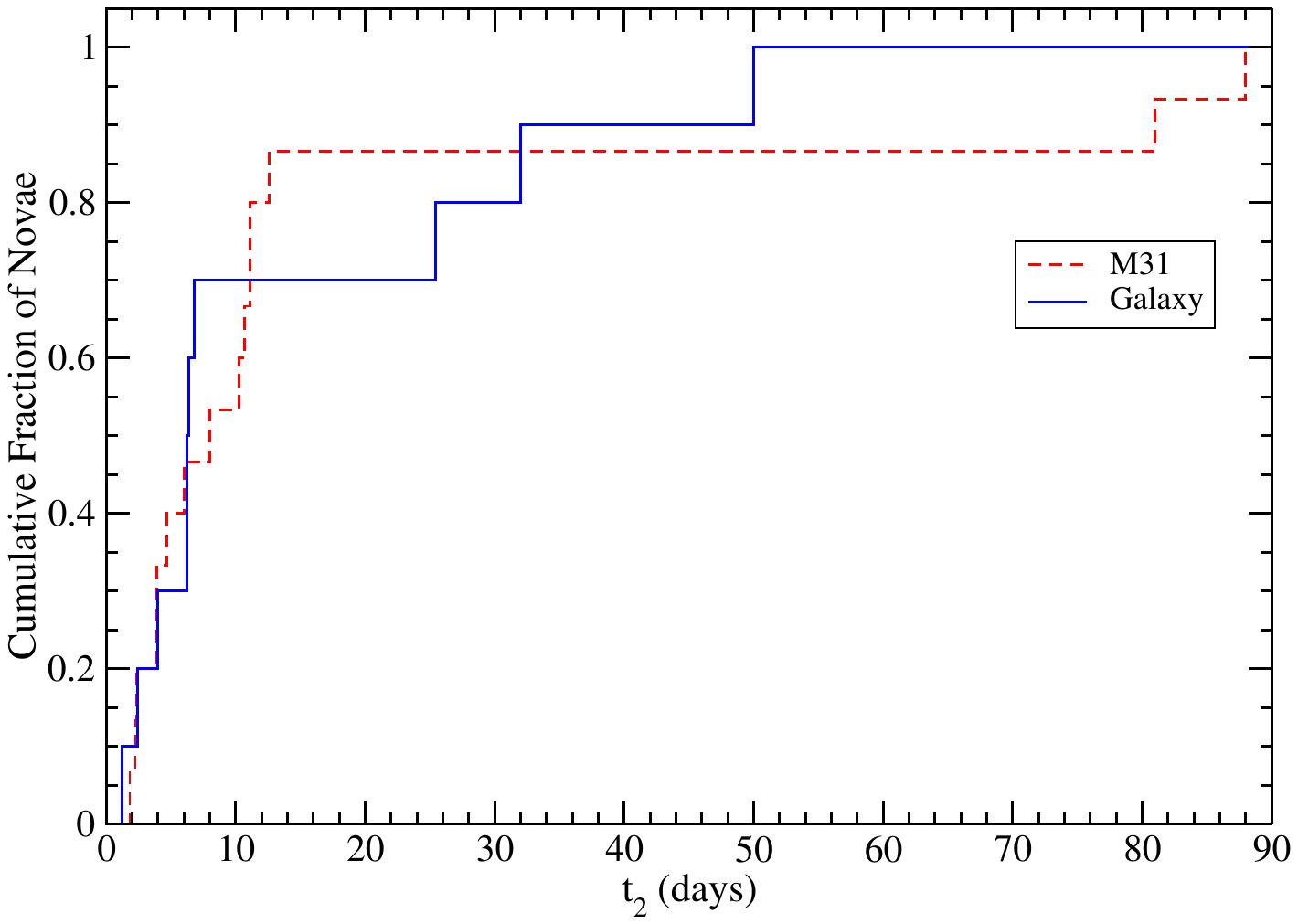}
\includegraphics[angle=0,scale=0.26]{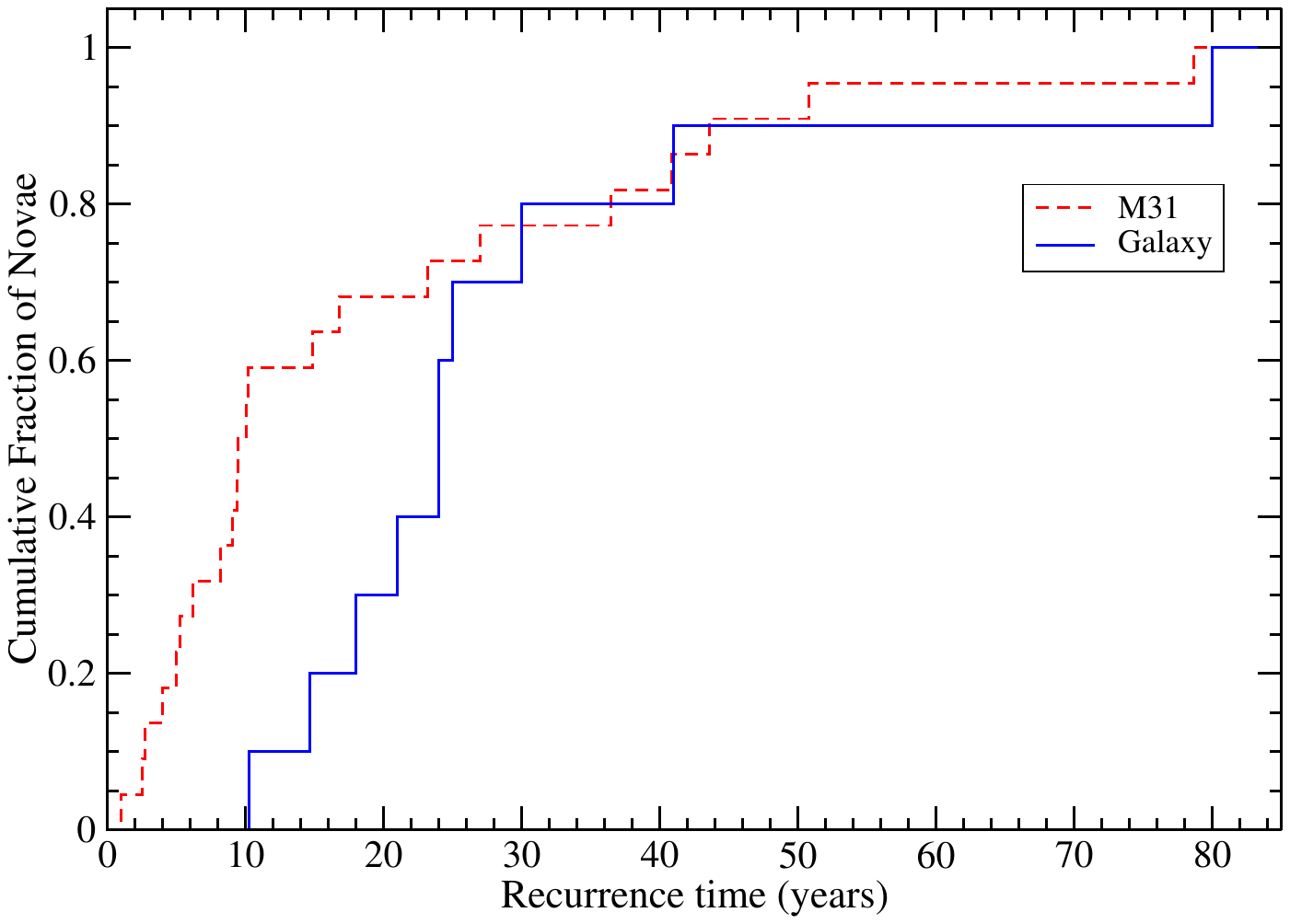}
\caption{
Left: The cumulative distributions of the $t_2$ times
for M31 RNe compared with Galactic RNe. There is no significant
difference between the two distributions (K-S $p=0.91$).
Right: The cumulative distributions of the recurrence
times for the M31 and Galactic RNe. A K-S test reveals that the M31 RNe
likely differ from their Galactic counterparts with K-S $p=0.07$.
}
\label{fig9}
\end{figure}

Clearly, small-number statistics make it difficult to draw
any firm conclusions from the comparison of the Galactic
and M31 RN populations.
Nevertheless, we compare the observed $t_2$ times and
recurrence times for the two galaxies in Figure~\ref{fig8}, along with their
cumulative distributions in Figure~\ref{fig9}.
There is no significant difference in the distribution of decline rates
between the two RN populations. The RN are overwhelmingly ``fast" novae
with typical decline rates $t_2\lessim10$~d. Interestingly,
although the vast majority of RNe are members of the He/N spectroscopic
class, the few outliers with $t_2>40$~d and available spectroscopy
(two in M31 and one in the Galaxy), show all three are members of the Fe~II
class.

Unlike the similar $t_2$ distributions, the recurrence-time distributions
for M31 and Galactic RNe appear to differ markedly. In particular,
while the shortest recurrence time known for any Galactic RN
is that of U~Sco with $T_\mathrm{rec}\simeq10.3$~yr \citep{Schaefer2010},
half of the known
M31 RNe have observed recurrence times of less than 10 years.
Even when the small number statistics are taken into account,
a K-S test shows that the two distributions would differ by more than
observed only 7\% of the time if they were derived
from the same parent distribution. So the observed difference is
intriguing.
It is unclear what observational selection effects might account for
the observed difference. In considering biases, it is important to
remember that the recurrence times are formally {\it upper limits\/}
on the true recurrence times. It is likely that outbursts of known
RNe have been missed, although it is not clear whether this bias
would differ between surveys of M31 and the Galaxy.

It is tempting to speculate that the difference in the recurrence time
distributions may involve the stellar populations of novae.
If the M31 observations are dominated by bulge
novae (given the more frequent sampling there), while the Galactic
sample is dominated by relatively nearby disk novae, it would
suggest that the bulge population
of RNe novae have shorter recurrence times on average compared with
disk RNe. Some support for this picture comes from
Schaefer \citep{Schaefer2022} who finds that novae with 
red giant secondaries are strongly associated with the Galactic bulge.
If we assume that such systems have higher mass transfer rates
than novae with unevolved secondaries, then for a given WD
mass, we would
expect the recurrence times of bulge novae to be shorter on average
compared with disk systems. Unfortunately, this explanation is in
conflict with
Williams et al. \citep{Williams2016} who found that novae with
red giant secondaries come primarily from the disk population of M31,
not the bulge. Thus, at present, we have no satisfactory explanation
for the observed difference in recurrence times between the RNe
populations in M31 and the Galaxy.

\section{Conclusions}

Our principal conclusions can be summarized as follows:

$\bf{(i)}$
After updating the work of Shafter et al. \citep{Shafter2015},
a total of 79 nova eruptions have now been detected in M31
associated with 20 firmly-established RNe,
with another four eruptions from two possible RNe.

$\bf{(ii)}$
The spatial distribution of known M31 RNe is indistinguishable from
that of novae generally, with the cumulative distributions of
both generally following the overall integrated light of the galaxy.

$\bf{(iii)}$
The peak luminosities and rates of decline for the M31 RNe
population differ markedly from novae generally. Most non-recurrent
novae broadly follow a MMRD relationship. On the other hand, the RNe population
is overwhelmingly concentrated in the lower-left quadrant
of the MMRD diagram, and is characterized by low peak luminosities
($M_V\grtsim-8$) and rapid rates of decline ($t_2\lessim10$~d).

$\bf{(iv)}$
The M31 and Galactic RNe populations are quite similar in terms
of their average peak luminosities and rates of decline, however
the recurrence time distributions appear to differ. Specifically,
half of the 22 M31 RNe have observed recurrence times that are
shorter than the shortest recurrence time known for a Galactic
RN (U~Sco with $T_\mathrm{rec} = 10.3$~yr).
A K-S test yields a $p$-value of 0.07,
providing marginal evidence that the difference between the two samples is real.
However, the cause of the observed discrepancy remains unclear.
Unknown and differing selection effects influencing the observations
of novae in the two galaxies may be at play,
and additional data will be needed before any firm conclusions can be drawn.

$\bf{(v)}$
Looking ahead, ongoing wide-field surveys
(e.g., ATLAS, ZTF, and especially the Vera C. Rubin Observatory,
which is now coming online) will continue to increase the volume
of high-cadence, near-continuous nova observations.
In addition,
the availability of new IR surveys made possible by facilities such as
WINTER, the Palomar observatory's Wide-field Infrared Transient Explorer
\citep{Frostig2024}
and the upcoming Nancy Grace Roman Space Telescope
will enable novae to be discovered in dust shrouded regions
of the Galaxy. Taken together, these current and planned surveys
promise to produce data that will substantially refine our understanding of
the nova recurrence-time distribution and the underlying WD mass
distribution on which it depends, both in the Milky Way and
across galaxies with widely varying stellar populations.

Key future efforts should focus on two priorities:
identifying and long-term monitoring of faint, fast novae
(which are likely recurrent but have been underrepresented in previous surveys)
and systematic surveys of galaxies spanning a broad range of Hubble types
to explore the sensitivity of nova properties to the underlying
stellar population.
These efforts will be particularly valuable if combined with
rapid multi-wavelength and spectroscopic follow-up
to better characterize individual systems
and test the potential role of RNe as progenitors of Type Ia supernovae.

\bibliographystyle{JHEP}
\bibliography{shafter}{}

\bigskip
\bigskip
\bigskip
\noindent {\bf DISCUSSION}

\bigskip
\noindent {\bf Massimo DELLA VALLE:} 
A simple comment about RNe as outliers of the MMRD.
The MMRD is tight (0.15 -- 0.2 mag) only when ejecta
masses are $10^{-4}$ -- $10^{-5}~M_\odot$ typical of CNe.
RNe have much lighter ejecta ($10^{-6}$ -- $10^{-7}~M_\odot$)
that become optically thin very quickly so they decline
in hours to a few days, and have dimmer maxima, therefore
they don't follow the MMRD.

\bigskip
\noindent {\bf Allen SHAFTER:}
The short recurrence times characteristic of RNe require progenitors
with high mass WDs accreting at high rates. Such systems
trigger a TNR with relatively small ignition masses, and so presumably
have small ejecta masses. And I agree with your basic point that the
low mass ejecta characteristic of RNe going optically thin quickly will
result in a relatively rapid decline. I'd also add that the high
accretion rates and short recurrence times of RNe keep the accreted layer
from becoming maximally degenerate prior to the TNR. So the peak
luminosity of the outburst will be generally lower than in typical CNe.

\bigskip
\bigskip
\noindent {\bf Ulisse MUNARI:} 
How much of the spread in the MMRD relation could be ascribed to
missed true maxima? At the time of plate surveys, they did not
observe for $\sim10$ nights each lunation around bright moon
(like asking how many points in the plot are from the
photographic era).

\bigskip
\noindent {\bf Allen SHAFTER:} 
You raise a good point. There will be some spread resulting
from missed maxima, even for the modern CCD surveys, however
none of the data shown in the MMRD plot is from the
older photographic surveys.

\end{document}